\documentclass{emulateapj}

\usepackage{graphics,graphicx}
\usepackage{natbib}
\def\apj{\rm ApJ}
\def\apjl{\rm ApJL}
\def\apjs{\rm ApJS}
\def\aj{\rm AJ}
\def\mnras{\rm MNRAS}
\def\nat{\rm Nature}

\def\aap{\rm AAP}
\newcommand{\msun}{M$_{\odot}$}

\shorttitle{Mergers, Flows, and Black Hole Growth}
\shortauthors{Bellovary et al.}

\begin{document}

\title{The Relative Role of Galaxy Mergers and Cosmic Flows in Feeding Black Holes}

\author{Jillian Bellovary\altaffilmark{1}, Alyson
Brooks\altaffilmark{2}, Marta Volonteri \altaffilmark{3}, Fabio
Governato\altaffilmark{4}, Thomas
Quinn\altaffilmark{4}, James Wadsley\altaffilmark{5}}

\altaffiltext{1}{Department of Physics and Astronomy, Vanderbilt University, Nashville, TN }

\altaffiltext{2}{Department of Astronomy, University of Wisconsin-Madison, Madison, WI}

\altaffiltext{3}{IAP, Paris, France}

\altaffiltext{4}{Department of Astronomy, University of Washington,
Seattle, WA}

\altaffiltext{5}{Department of Physics and Astronomy, McMaster
University, Hamilton, ON, Canada}

\begin{abstract}

\noindent

Using a set of zoomed-in cosmological simulations of high-redshift
progenitors of massive galaxies, we isolate and trace the history of
gas that is accreted by central supermassive black holes.  We
determine the origins of the accreted gas, in terms of whether it
entered the galaxy during a merger event or was smoothly accreted.
Furthermore, we designate whether the smoothly accreted gas is
accreted via a cold flow or is shocked upon entry into the halo.  For
moderate-mass ($10^6 - 10^7$ \msun) black holes at $z \sim 4$, there
is a preference to accrete cold flow gas than gas of shocked or merger
origin.  However, this result is a consequence of the fact that the
entire galaxy has a higher fraction of gas from cold flows.  In
general, each black hole tends to accrete the same fractions of
smooth- and merger-accreted gas as is contained in its host galaxy,
suggesting that once gas enters a halo it becomes well-mixed, and its
origins are erased.  We find that the angular momentum of the gas upon
halo entry is a more important factor; black holes preferentially
accrete gas that had low angular momentum when it entered the galaxy,
regardless of whether it was accreted smoothly or through mergers.

\end{abstract}


\section{Introduction}

The classic model for fueling bright quasars at high redshift assumes
that massive black holes (MBHs) are powered by major galaxy mergers,
which cause large quantities of gas to be funneled into the regions
where MBHs reside
\citep[e.g.][]{Kauffmann00,Haiman01,DiMatteo05,Hopkins06,Li2007,Sijacki09}.
The result of this process is near-simultaneous bursts of star
formation and MBH accretion, which are observed at all redshifts in a
large range of galaxy types.  Indeed, the peaks in global star
formation and quasar activity coincide at a redshift of $\sim 2-3$
\citep{Hopkins04,Richards06,Reddy08}, which strengthens the argument
that these two phenomena must be connected.  Observations support this
picture, with growing numbers of observed high-redshift galaxies
exhibiting both powerful star formation and active galactic nucleus
(AGN) activity
\citep{Alexander05,Papovich06,Menendez-Delmestre07,Silverman09,Coppin10,Mullaney12,Kirkpatrick12,Chen13}.
Nearby massive early-type quasar host galaxies are also observed to
exhibit signs of a recent major merger
\citep{Sanders88,Disney95,Bahcall97,Canalizo01,Hutchings03,Guyon06,
  Bennert08,Urrutia08,Wolf08,Tal09,Schawinski10}.  A popular suggested
paradigm involves a gas-rich major merger triggering a burst of star
formation, channeling gas into the central regions of the galaxy, and
fueling a powerful AGN.  The energy released by accreting gas may
unbind the remaining gas from the galaxy, rendering the galaxy ``red
and dead'' and eventually resembling today's massive ellipticals
\citep{DiMatteo05,Hopkins06}.

However, there is also evidence for AGN activity that is not necessarily
merger-driven.  Low to moderate luminosity AGN hosts tend to be
disk-dominated at high redshifts \citep{Schawinski11,Kocevski12} and
in the local universe \citep[e.g.][]{DeRobertis98}.  Recently,
\citet{Treister12} postulated that major mergers are the primary
drivers of accretion in only the most luminous AGN.  Furthermore, they
suggest that this trend does not appear to evolve with redshift;
moderate-luminosity AGN inhabit quiescent galaxies out to $z \sim
2-3$.

Locally, low-luminosity AGN hosts tend to be disk galaxies, which
are unlikely to have undergone a recent major merger.  Recent studies
have examined MBH fueling in local AGN and found that gravitational
torques are efficient at transporting gas from the galactic disk to
the central region \citep{Haan09,Garcia-Burillo09}.  \citet{Ryan07}
examined a sample of narrow-line Seyfert 1 galaxies and found no
evidence for merger events, tidal tails, or disrupted features. The
lack of merger-induced features for these galaxies is a hint that
their black hole growth is driven by secular processes rather than
interactions with other galaxies.  Furthermore, \citet{Hicks13}
demonstrate that in a matched sample of Seyfert and quiescent
galaxies, AGN hosts are more centrally concentrated with a larger
reservoir or molecular gas, perhaps indicating that active nuclei have
simply experienced recently triggered inflowing material.

In addition to secular processes, and especially at high redshifts,
MBHs may be fueled by gas which is directly accreted onto a galaxy
from the intergalactic medium.  These ``cold flows'' may take the form
of filamentary inflows, especially at high redshift
\citep{Binney77,Keres05,Dekel06,Ocvirk08}.  Such flows fueling MBH
growth have been demonstrated in large volume and high resolution
cosmological simulations \citep[][respectively]{DiMatteo12, Dubois12}.
The role of cold flows on MBH growth may be significant among the most
massive rare halos, but the effect on more common, less massive halos
has yet to be investigated, and is the primary goal of this paper.

The diversity of AGN hosts suggests that there exists a diversity of
fueling mechanisms.  For instance, while not all AGN activity is
merger driven, mergers or interactions seem to enhance AGN activity
\citep{Silverman11}.  Properties such as galaxy/halo mass, merger
history, gas accretion history, and environment likely all play a role
in when and how much MBHs accrete.  Thus, the aforementioned paradigm
of major mergers fueling MBH growth is likely only a dominant
mechanism for a small subset of highly luminous objects.  The question
``how do MBHs grow?'' is likely a very complex one for black holes in
different environments.  To address one aspect of this question, we
focus on late-type galaxies at $z = 4$ which host moderate-luminosity
AGN.  The advantage of studying MBH growth in cosmological simulations
is that one can disentangle effects from major galaxy mergers as well
as accretion of ambient cold gas from cold flows/filaments, minor
mergers, and secular evolution.

In this work we separate accretion by mergers and the ambient
environment by tracing the history of every gas particle as it enters
the primary halo.  Furthermore, we differentiate between cold flows
and gas which is shocked upon entering the virial radius of the halo.
We then trace each gas particle on its way to being accreted by the
central MBH.  In section 2 we describe the simulations used in our
study.  Section 3 describes the method used to trace the history of
the gas particles.  In section 4 we reveal the role of the origins of
gas as well as its angular momentum in fueling central MBHs.  We leave
studies of minor mergers and secular processes to a future work, and
summarize our results in section 5.

\section{The Simulations}

We have run our cosmological simulations using the Smoothed Particle
Hydrodynamics $N$-Body Tree code Gasoline \citep{Stadel01,Wadsley04}.
Gasoline includes prescriptions for star formation and supernova
feedback \citep{Stinson06}, metal diffusion \citep{Shen10}, and
massive black hole formation, accretion, and feedback
\citep{Bellovary10}. All of our simulations use a Kroupa IMF
\citep{Kroupa}, WMAP3 cosmology \citep{WMAP3}, and a uniform UV
background \citep{Haardt96}.  Galaxies simulated with Gasoline have
been shown to lie on the observed Tully-Fisher relation
\citep{Governato09}, the mass-metallicity relation \citep{Brooks07},
the size-luminosity relation \citep{Brooks11}, and have realistic
baryon fractions and matter distributions
\citep{Governato10,Guedes11}.  Our simulations use primordial cooling
plus a low-temperature extension to the cooling curve due to trace
metals \citep{Bromm01}.  While we do not include cooling via metals or
molecular hydrogen, gas is allowed to reach a minimum temperature of
$\sim 100$K.  The abovementioned works demonstrate that we are able to
simulate realistic galaxies without the addition of these advanced
features \citep[see also][]{Christensen13}.  While these previous
results have not included the effects of AGN feedback, our current
studies suggest that for galaxies of small to moderate masses (such as
those studied previously and in this work) the central SMBH affects
only the central regions of the galaxies and leaves the more global
properties intact; we are thus confident that the simulations
presented in this paper are quite consistent with the broad range of
observed galaxy properties.

We have selected three halos of interest from a uniform volume of size
50 Mpc, and re-simulated each of them individually using the volume
renormalization technique \citep{Katz92}.  The selected galaxies are
the progenitors of today's massive ellipticals, with $z = 0$ virial masses of
between $1-9 \times 10^{13}$; however, we have chosen to maximize
resolution and evolve these galaxies only until redshift $z \sim 4$.
The force resolution is 260 comoving pc, and the gas and dark matter
particles masses are $9 \times 10^4$ \msun~ and $1.3 \times 10^5$
\msun~ respectively.  Table \ref{table:getgasgals} describes the
properties of the simulations at $z = 4$ in detail, including the
number of particles within the virial radius, the virial mass, maximum
value of the circular velocity, virial radius (defined as the radius
in which the enclosed density is 200 times that of the critical
density of the universe at z=4), and stellar disk scale radius as
measured with an exponential profile.

\begin{deluxetable*}{lcccccllc}
\tablecolumns{8} 
\tablewidth{0pc}
\tablecaption{Simulation Properties\label{table:getgasgals}}

\tablehead{
\colhead{Run} & \colhead{\# within } &\colhead{M$_{vir}$ } &\colhead{M$_{gas}$ }  &\colhead{M$_{star}$ }  &\colhead{V$_{max}$ } & \colhead{R$_{vir}$} & \colhead{$r_s$} \\
\colhead{}&\colhead{R$_{vir}$}&\colhead{(\msun)}&\colhead{(\msun)}&\colhead{(\msun)}&\colhead{(km/s)}&\colhead{(kpc)}&\colhead{(kpc)} }

\startdata
\hline

hz2 & 2758081 & $2.52 \times 10^{11}$ & $2.36 \times 10^{10}$& $1.63 \times 10^{10}$& 267 & 196 & 2.39 \\
hz3 & 3516774 & $3.17 \times 10^{11}$ &  $3.62 \times 10^{10}$& $2.06 \times 10^{10}$& 190 & 211 & 2.51 \\
hz4 & 527952 & $5.81 \times 10^{10}$ &  $7.53 \times 10^{9}$&$9.64 \times 10^{8}$ & 106 & 120 & 0.80  \\

\enddata
\end{deluxetable*}

\begin{figure}[htb!]
  \begin{center}
  \includegraphics[scale=0.5]{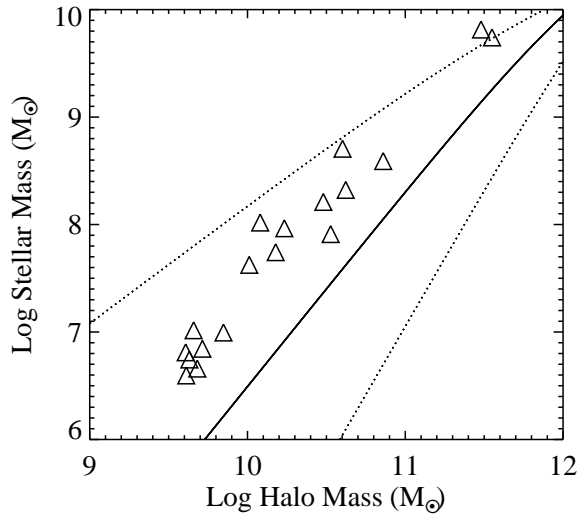}
\caption{ 
   \label{fig:moster}  
   The stellar mass - halo mass relation for every single galaxy with
   more than 20,000 dark matter particles in all three simulations at
   $z = 4$. The solid line is the analytical relation derived for $z =
   3.5$ and the dotted lines encompass the 1-$\sigma$
   errors for that relation \citep{Moster10}.  }
    \end{center}
\end{figure}

Star formation is modeled stochastically, with gas particles becoming
eligible to form stars if they are above the density threshold (2.5
amu cm$^{-3}$) and below the temperature threshold (10$^4$ K).  We use
a star formation efficiency parameter of $c^* = 0.1$, and a supernova
feedback energy of $E_{SN} = 8 \times 10^{50}$ erg.  This energy is
distributed to particles within a radius determined by the blastwave
equations of \citet{McKee77}; the heated particles have their cooling
disabled for a timescale also determined by these equations.
This combination of parameter choice and resolution results in
galaxies which are consistent with the stellar mass - halo mass
relation \citep{Moster10,Munshi12} at $z \sim 4$ (Figure
\ref{fig:moster}) and have realistic star formation histories.  Thus,
while there are few observational structural constraints at $z=4$ for
late-type galaxies, we are confident that we can adequately represent
galaxy and MBH growth at this epoch.

Seed black hole formation is also stochastic, and is similar to star
formation with the additional requirement that black holes must form
out of zero-metallicity gas.  If a gas particle meets the criteria for
star formation and additionally has zero metallicity, it has an
additional probability, $\chi_{seed}$, that it will instead form a
black hole with mass equal to the mass of the parent gas particle ($9 \times 10^4$) \msun~.
For our simulations here we set $\chi_{seed} = 0.2$, which achieves
reasonable MBH occupation fractions in galaxies without overproducing
MBH-MBH merger events (which results in MBHs which violate the Soltan
argument, growing almost solely through mergers and not gas
accretion).
This requirement is broadly consistent with theories of seed black hole
formation out of Population III stars
\citep{Couchman86,Abel02,Bromm04} or direct collapse of gas
\citep{Loeb94,Eisenstein95,Koushiappas04,Begelman06,Lodato06}.  It
also ensures that seed black holes cease forming at high redshift,
once the interstellar medium becomes polluted with metals from
supernova feedback.  Multiple black holes may form in a single halo;
additionally, they are not fixed to halo centers but are allowed to
react dynamically to perturbations such as galaxy mergers.  In order
to decrease the effects of artificial two-body scattering, we employ
dark matter particle masses that are only slightly larger than the
black hole masses, which helps the black holes stay in the centers of
their hosts.  For more details on seed black hole formation, see
\citet{Bellovary11}.

\begin{figure}[htb!]
  \begin{center}
  \includegraphics[scale=0.5]{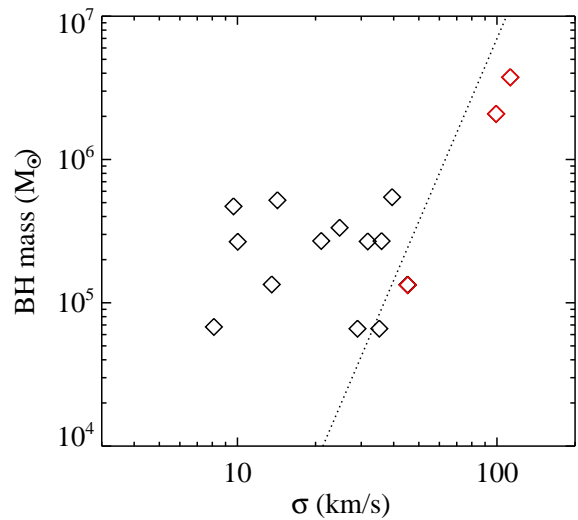}
\caption{ 
   \label{fig:msigma}  
   The $z = 4$ M-$\sigma$ relation for galaxies in all three
   simulation boxes.  The three central galaxies are highlighted in
   red.  The dotted line is the local relation derived by
   \citet{Gultekin09}.  The ``plume" of massive objects at low $\sigma$ is
   a result of our relatively massive MBH seeds 
   \citep[see][]{Volonteri09}.}
    \end{center}
\end{figure}

Black holes are allowed to merge if they {\em (a)} are within twice
one another's softening length and {\em (b)} fulfill the criterion
$\frac{1}{2} \Delta \vec{v}^2 < \Delta \vec{a} \cdot \Delta \vec{r}$,
where $\Delta \vec{v}$ and $\Delta \vec{a}$ are the differences in
velocity and acceleration of the two MBHs, and $\Delta \vec{r}$ is the
distance between them.  Black holes also grow through gas accretion,
via the Bondi-Hoyle method:

\begin{equation}
\dot M = \frac{4\pi \alpha G^2M_{BH}^2\rho}{(c_s^2 + v^2)^{3/2}}
\end{equation}

\noindent
where $\rho$ is the density of the nearby gas, $c_s$ is the sound
speed, $v$ is the relative velocity of the black hole to the gas, and
$\alpha$ is a constant equal to 1.  Feedback energy is imparted on the
surrounding gas and is proportional to the accretion rate:

\begin{equation}
\dot E = \epsilon_r \epsilon_f\dot M c^2
\end{equation}

\noindent
assuming $\epsilon_r = 0.1$ as the radiative efficiency, and
$\epsilon_f = 0.03$ as the efficiency at which the feedback energy,
which is distributed over the SPH kernel as thermal energy to the 32
particles nearest the black hole, couples to the gas.  Other groups
often use a value of $\epsilon_f = 0.05$
\citep[e.g.][]{Sijacki07,DiMatteo08}; however, we find that using this
higher value results in overpowered feedback and very limited gas
accretion.  Using $\epsilon_f = 0.03$ results in MBHs which match the
observed MBH-host galaxy scaling relations, such as $M_{BH}-\sigma$
and $M_{BH}-M_{bulge}$ (e.g. Figure \ref{fig:msigma}).  Since this number
is not meant to represent a precise physical quantity, but is merely a
parameter used in subgrid feedback models in simulations, slight
differences among various codes are expected.  In addition, in this
work we are examining the relative contribution of various channels of
gas accretion rather than an absolute value of accreted gas mass, and
thus our results are not highly dependent on our feedback
implementation.

\section{Tracing Gas Accretion}

The following section is a summary of the methodology described in
\citet{Brooks09}, which is a comprehensive study of how galaxies
accrete gas in cosmological simulations.  While the focus of
\citet{Brooks09} is on the gaseous and stellar buildup of galaxy
disks, the gas-tracing methodology can be applied to many scenarios,
including the accretion of gas by black holes.

Galaxies are identified using the Amiga Halo Finder
\citep[AHF;][]{Knebe01,Gill04,Knollmann09}, which identifies a virial radius
based on the overdensity criterion for a flat universe
\citep{Gross97}. Before their accretion onto the primary halo, gas
particles are traced to determine if they ever belonged to a halo
other than the primary.  To define the primary halo, we identify the
central MBH of the most massive galaxy at the final step of the
simulation.  We then trace this MBH back in time and identify its host
as the primary at each step.

 If gas particles are already in the primary halo at the
first step where halos are identified ($z = 15$ in this study), they
are labeled as ``early'' accretion.  Gas particles that have ever
belonged to any halo or subhalo other than the primary halo are
labeled as ``clumpy'' accretion, and are generally accreted onto the
main galaxy in merger events.  Gas that enters the halo in any other
fashion is labeled ``smooth'' accretion, which is then divided into
two categories, cold and shocked.

Smoothly accreted gas particles may or may not undergo a shock within
the virial radius of the primary galaxy
\citep{Dekel06,Keres09,Vandevoort11}.  The presence of a shock depends
mainly on the mass of the halo; generally, gas will shock if the
galaxy halo is around $10^{11}$ \msun \citep{Ocvirk08}.  Thus, since
dwarf galaxies are less massive than this threshold, they will never
develop a shock.  In massive galaxies, the development of a shock is
strongly related to the orientation of the gas flow.  Cold gas
predominantly flows into halos along the filaments inherent to the
large-scale structure of the universe, and these filaments may be
dense enough to penetrate a shock front and deliver gas to the central
galaxy without heating the gas to the virial temperature
\citep{Keres09,Dekel09,Brooks09,Stewart11b,Vandevoort11,Stewart13}.
Eventually the filaments dissipate, but for a substantial time these
filaments are critical in building up galaxy disks at high redshift
\citep{Brooks09,Pichon11,Danovich12}.

Shocked gas by definition undergoes an entropy increase, which also
manifests as a temperature increase.  Thus, gas is identified as
shocked if it increases its entropy and density based on these two
criteria \citep[see also][]{Dekel06}:

\begin{equation}
T_{shock} \geq 3/8T_{vir},
\end{equation}

 where $T_{vir}$ is the virial
temperature of the halo, and 

\begin{equation}
\Delta S \geq S_{shock} - S_0,
\end{equation}

 where $S_0$ is the initial entropy of the gas particle and 
 
 \begin{equation}
S_{shock} = log_{10}[3/8~T_{vir}^{1.5}/4\rho_0]
\end{equation}

\noindent
where $\rho_0$ is the gas density prior to encountering the shock.
Smoothly accreted gas is traced until it reaches a distance of
$0.1R_{vir}$ from the galaxy; after this point, supernova feedback can
mimic the effects of virial shocking, which can no longer be
accurately tracked.  Any smoothly accreted gas particle that undergoes
the above temperature and entropy increases is defined as shocked.
Those gas particles that reach $0.1R_{vir}$ without these increases is
defined as unshocked (repeating the calculation for tracking radii
larger than $0.1R_{vir}$ gives no substantial difference in our
results).  We refer the reader to \citet{Brooks09} for more complete
details of the relevant shock physics.

Armed with these categories of gas accretion, we determine how MBHs
accrete their own gas.  We identify the gas particles which
are 
accreted by the primary MBH in a simulation (defined as the MBH residing
at the center of the primary galaxy).  We then cross-correlate these
particles with their accretion history.  This process provides a
history for the accreted gas particles as they enter the primary
galaxy, though does not give any information about which mechanisms
govern their journey to the MBH.  This information must be gleaned in
further analysis.

\section{How Do Black Holes Get Their Gas?}

\begin{figure*}[htb!]
  \begin{center}
  \includegraphics[scale=0.8]{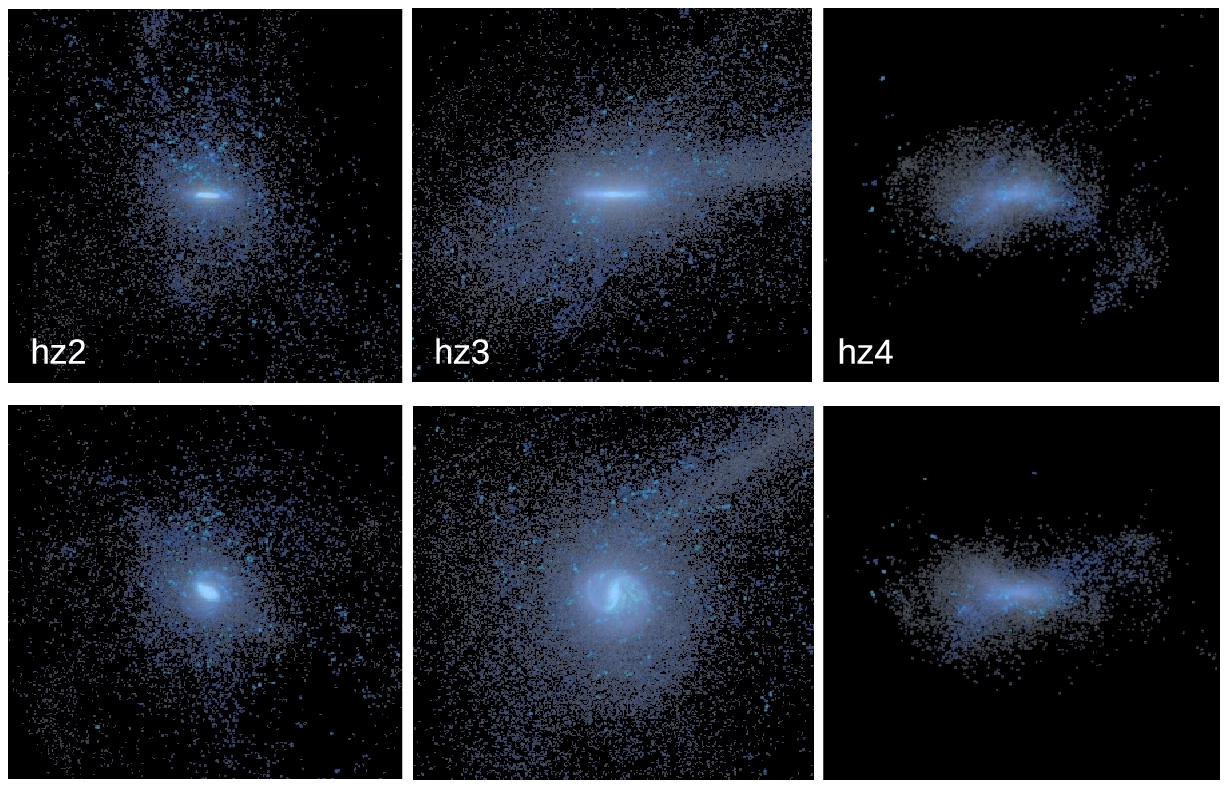}
\caption{ 
   \label{fig:images}  
   Edge-on (top row) and face-on (bottom row) of the primary galaxies
   in simulations $hz2$, $hz3$, and $hz4$ (left, center, and right,
   respectively).  Images are unreddened SDSS $gri$ composites created
   with Sunrise \citep{Jonsson06} and are 30 kpc on a side.  The rest-frame unreddened absolute magnitudes in the $Hubble$ ACS F435 band are -23.6, -24.0, and -20.9, respectively. }
    \end{center}
\end{figure*}

\begin{figure*}[htb!]
  \begin{center}
  \includegraphics[scale=0.8]{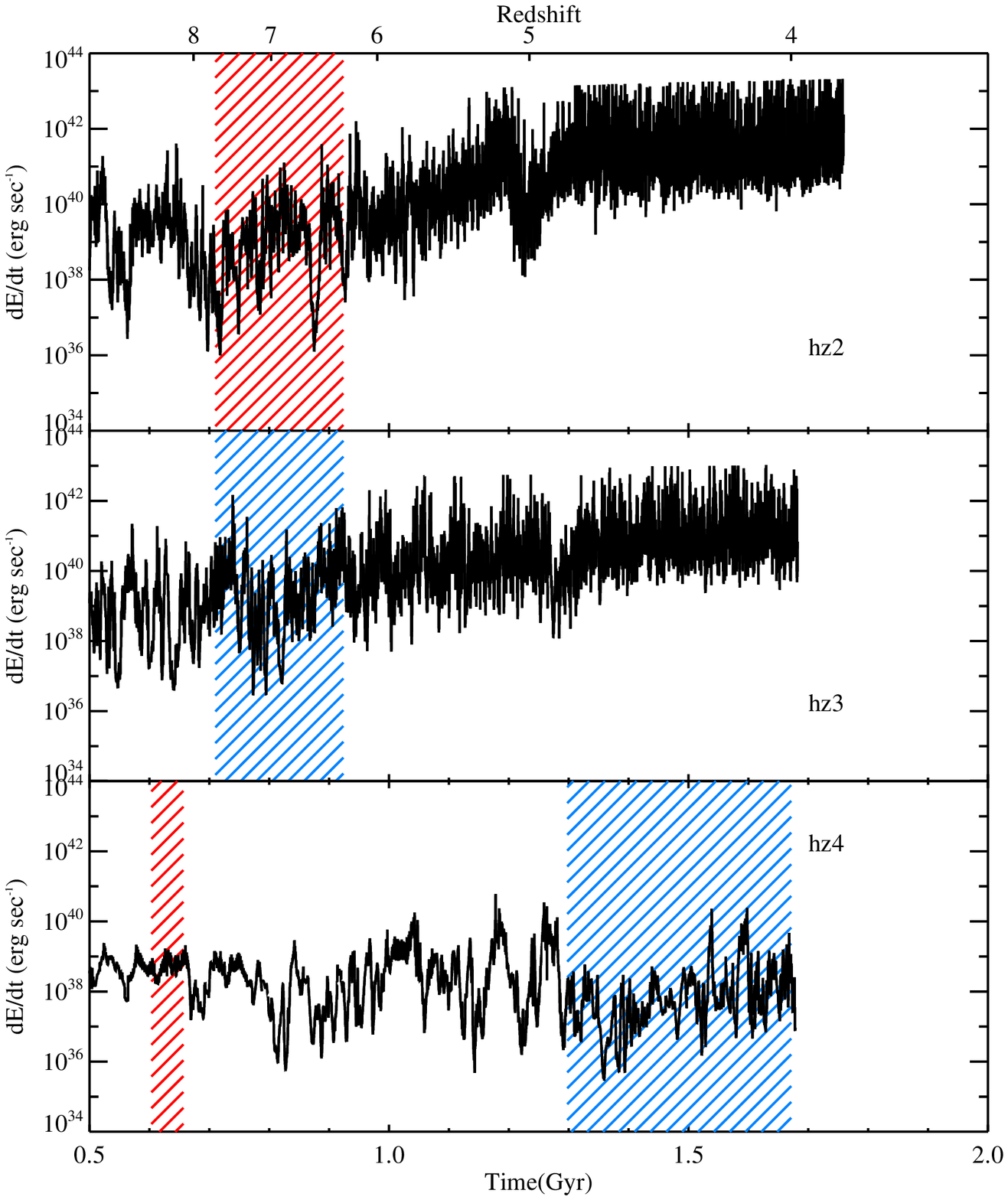}
\caption{ 
   \label{fig:lightcurves}  
   Bolometric luminosity from accretion onto the primary MBH vs time
   for the most massive galaxy in $hz2$ (top), $hz3$ (middle), and
   $hz4$ (bottom).  Major mergers are marked with a red hatched
   region, and minor mergers with a blue hatched region. Merger events
   do not substantially affect the growth of these lower mass MBHs.
   }
    \end{center}
\end{figure*}

Figure \ref{fig:images} shows simulated SDSS $gri$ images of the
primary galaxies in each simulation, which were created with Sunrise
\citep{Jonsson06}.  The images show blue colors and late-type
morphologies for all three galaxies.  Our three chosen simulations have
similar mass and environment.  They each form in gas-rich filaments
and accrete gas from this cold flow as well as through galaxy mergers.
However, they have a variety of merger histories; for example, the
largest galaxy ($hz2$) has an active history, undergoing a few major
mergers (defined as mergers with stellar mass ratio 1:3 or larger).
The galaxies $hz3$ and $hz4$ are somewhat more quiescent, with mostly
minor mergers (stellar mass ratio between 1:3 and 1:10).  We have
chosen these simulations in order to examine how galaxies with
different interaction histories may experience different MBH growth.
In Figure \ref{fig:lightcurves} we show the bolometric luminosity vs
time for the central MBH in the primary galaxy for each simulation.
We highlight major merger events with red hatched regions, and minor
mergers with blue hatching.  Mergers are defined to begin when the
satellite enters the primary's virial radius, and end when the
satellite is disrupted.  While each lightcurve is fairly noisy,
mergers do not seem to substantially enhance the fueling of low-mass
MBHs.  However, major mergers trigger strong AGN activity only after
the third pericenter passage, when central MBHs are $\sim 1$ kpc apart
\citep{VanWassenhove12}, and our resolution limits our ability to
analyze this phase in great detail. 
At lower redshifts there are several extremely minor mergers (with
ratios less than 1:10) as well as a variety of secular instabilities
which occur during each galaxy's evolution; the MBH appears to undergo
at least as many spurts of growth during these episodes as during more
substantial mergers.  The MBH in $hz2$ reaches a bolometric luminosity
greater than $10^{42}$ erg s$^{-1}$ for a duration of 66 Myr, or 4.5\%
of its lifetime, and exceeds a luminosity of $10^{43}$ erg s$^{-1}$
for a total of 3.7 Myr.  The MBH in $hz3$ spends 15 Myr (1\% of its
lifetime) at a luminosity greater than $10^{42}$ erg s$^{-1}$. The MBH
in $hz4$ experiences very little growth by accretion.  The
luminosities of the MBHs in $hz2$ and $hz3$ are comparable to those of
nearby Seyfert galaxies, but are unobservable at the high redshifts of
our simulations.

\begin{figure*}[htb!]
  \begin{center}
  \includegraphics[scale=0.35]{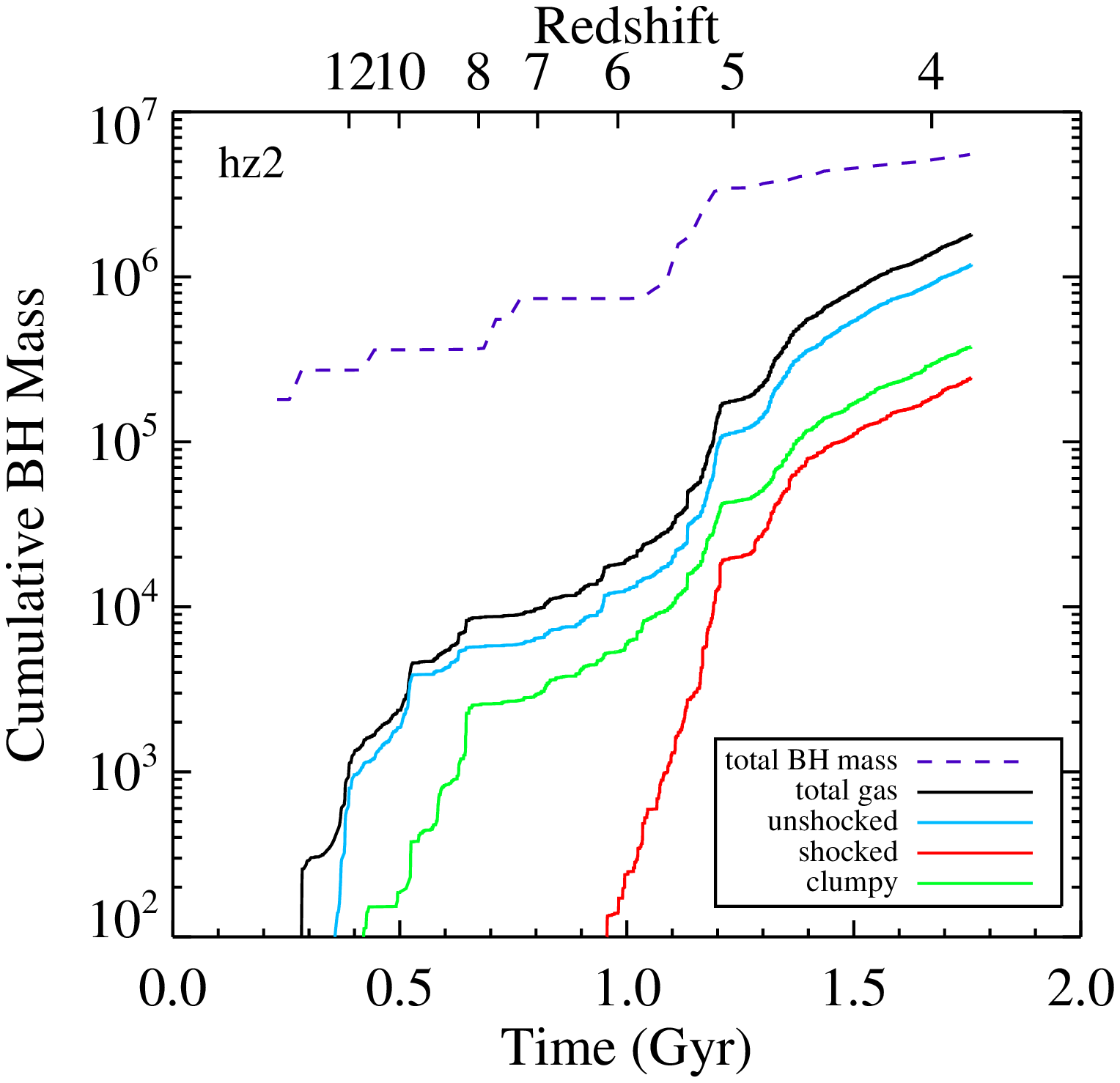}
  \includegraphics[scale=0.35]{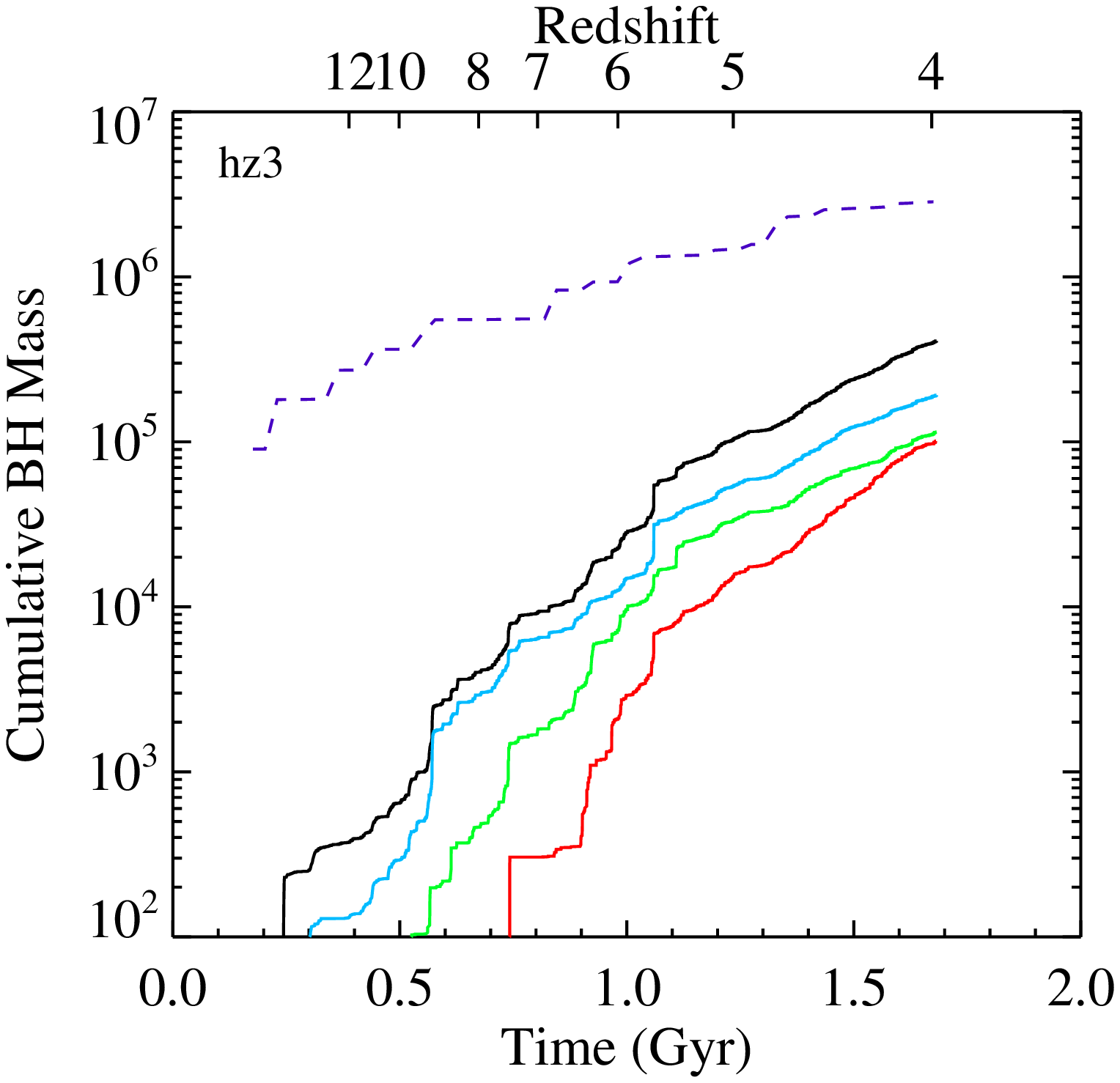}
  \includegraphics[scale=0.35]{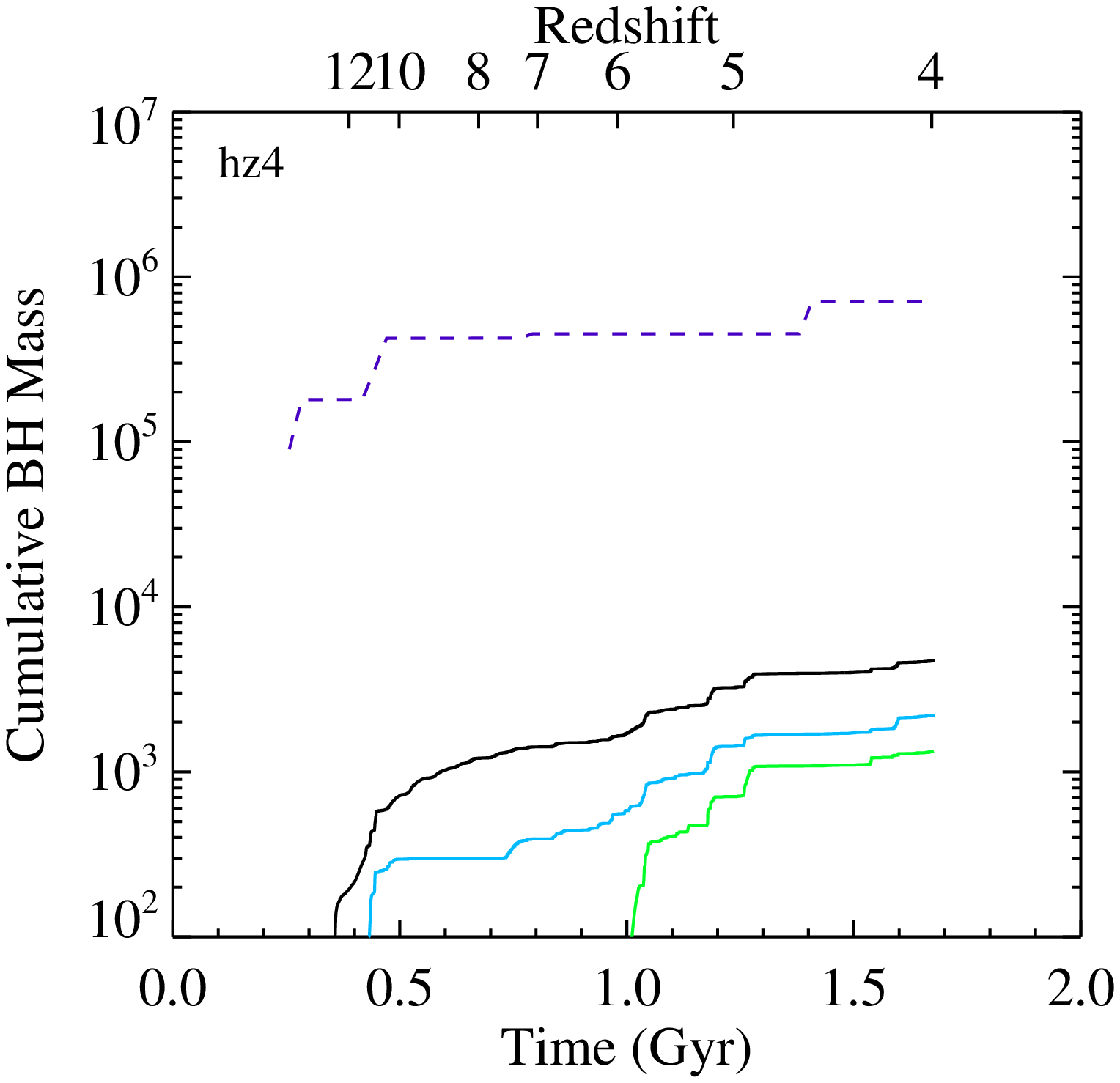}
\caption[Stuff]{ 
   \label{fig:BHgrowth}  
    Growth of the primary MBH in the the largest galaxy in simulation
  $hz2$ (left), $hz3$ (center), and $hz4$ (right).  The upper dashed
  line is the total mass of the MBH.  The solid black line is the
  total resulting from accreted gas.  The blue, green, and red lines
  represent accreted gas originating from unshocked flows, mergers, and
  shocked accretion, respectively.  }
    \end{center}
\end{figure*}

We show the growth of the primary black hole in the largest galaxy in
each run in Figure \ref{fig:BHgrowth}.  Each MBH begins with a seed
mass of $9 \times 10^4$ \msun, and subsequently grows through merging
with other MBHs and via gas accretion.  The purple dashed line
indicates the total MBH mass, including the initial seed mass and
growth through MBH mergers.  The solid black line represents the total
mass gained by gas accretion only.  The colored solid lines show the
mass due to gas accretion divided into each accretion mode:
unshocked/cold flows (blue), shocked (red), and clumpy (green).

In each galaxy, we see that gas which originated in cold flows 
dominates the MBH accretion. 
This result indicates that gas from other galaxies does not always
travel directly to the central MBH during mergers, but may get mixed
into the host galaxy's gas reservoir, or delayed in entering the
central regions due to an elongated orbit.  
Cold flow gas appears to be the dominant fuel for MBHs; we must
investigate whether this result is due to cold flows being more
efficient MBH feeders, or whether they simply provide the most gas to
their host galaxies.


\subsection{Gas Fractions}


\begin{deluxetable*}{l|ccc|ccc|ccc}
\tablecolumns{10} 
\tablewidth{0pc}
\tablecaption{Galaxy vs. MBH Accretion\label{table:gasfractions}}

\tablehead{  & & hz2 & & & hz3 & & & hz4 & \\
\colhead{Gas Origin} &\colhead{gas} & \colhead{stars}&\colhead{MBH} &\colhead{gas} & \colhead{stars} & \colhead{MBH} & \colhead{gas} & \colhead{stars}&\colhead{MBH}}

\startdata

Unshocked & 62\% &60\%& 66\%   & 50\% & 42\%&47\%  & 65\% & 58\% &47\%\\
Clumpy    & 25\% & 28\%&21\%   & 30\% & 35\%&28\%  & 30\% & 35\%&28\%\\
Shocked   & 13\% & 12\%&14\%   & 20\% & 23\%&25\%  & 0\%  & 0\%&0\%\\

\enddata
\end{deluxetable*}

While Figure \ref{fig:BHgrowth} gives insights on the mode of gas
which is accreted by MBHs, it does not give us any information on
whether one mode is preferentially accreted over another.  To do this,
we must compare the fraction of each mode accreted by the MBH to the
fraction of each mode that composes our entire galactic halo.  If a
halo is comprised of gas hailing from one particular mode of
accretion, it follows that the MBH will also accrete the majority of
its gas from the same mode.  On the other hand, if the modal fractions
accreted by the MBH are different from those comprising the overall
galaxy, we can clarify whether one mode is preferentially accreted
over another.

Table \ref{table:gasfractions} shows the fractions of unshocked,
clumpy, and shocked gas accreted by the galaxy and MBH for each of the
three simulations.  The ``stars" column is the fraction of stellar
mass which formed in-situ from gas originating in each accretion mode
(stars which formed in another galaxy are not included).  For
simulation $hz4$, the primary is never large enough to host a shock.
The remainder of the gas in $hz4$ is classified as ``early''
accretion: gas which was already in the galaxy at $z = 15$.  For $hz2$
and $hz3$ the early fraction is negligible.

Cold flows are generally the dominant source of accretion for both the
overall galaxy and the central MBH, followed by gas from mergers.
Overall, there is very little difference between the modal fractions
of accreted gas for the galaxies and the MBHs.  The stars, on the
other hand, seem to show a very slight relative preference to form
from merger gas compared to the MBH's preference for unshocked gas (in
relation to the overall galaxy composition).  While a detailed study
of this phenomenon is beyond the scope of this paper, we suspect that
gas entering the main galaxy during a merger may experience a variety
of processes which may cause it to form stars; on the other hand, to
feed the MBH it must be channeled to a specific location.
We stress that while these trends
are evident for all three galaxies in our sample, the differences are
only a few percent.  Overall, each black hole tends to accrete more or
less the same fractions of smooth- and merger-accreted gas as is
contained in its host galaxy.

\begin{figure}[htb]
  \begin{center}
    \includegraphics[scale=0.5]{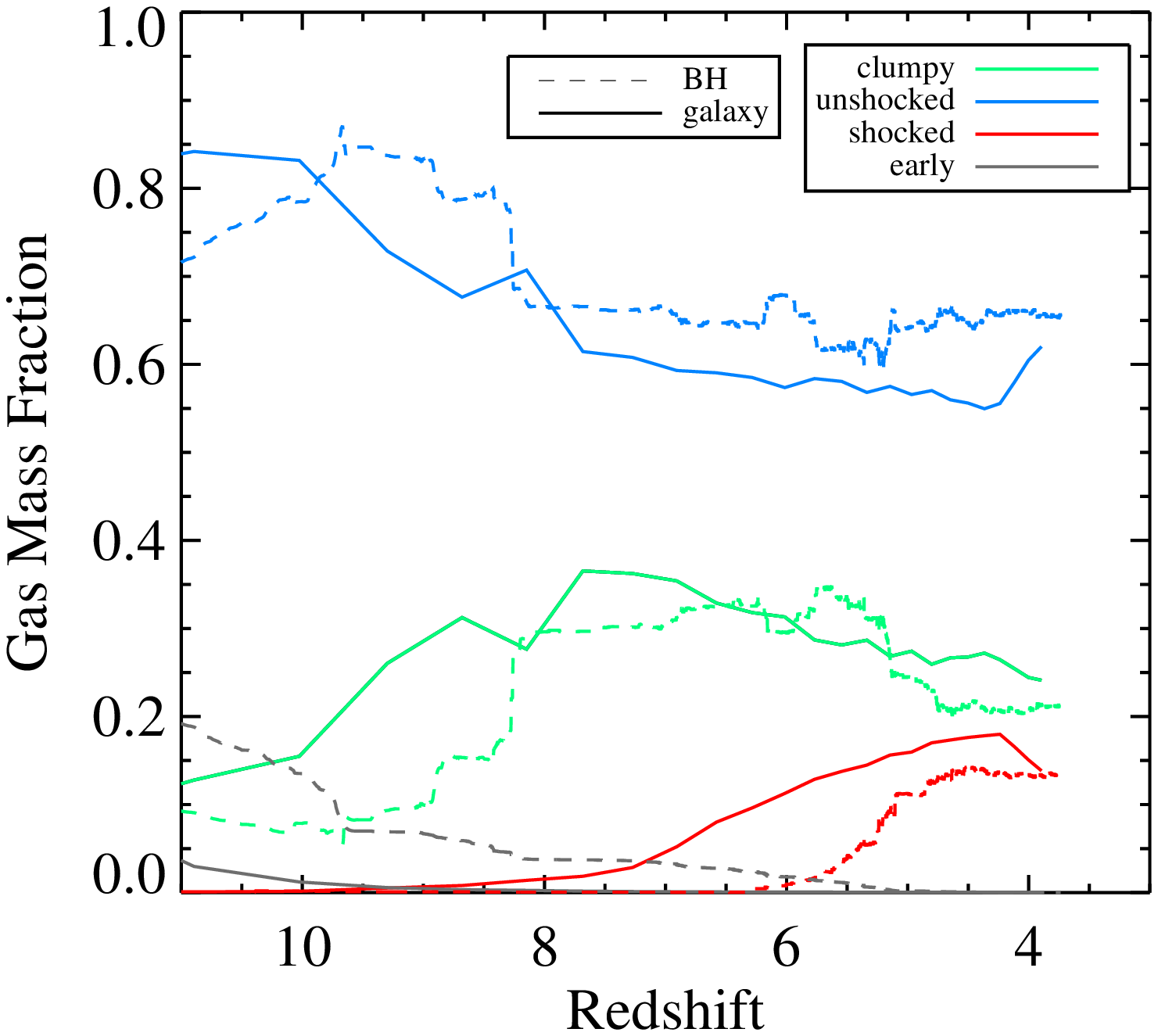}
\caption{ 
   \label{fig:gasfraction}  
   Cumulative gas fractions vs redshift for the primary galaxy in
   simulation $hz2$.  Colors are as in Figure \ref{fig:BHgrowth}, with
   the addition of a gray line representing gas which is within the
   galaxy at the beginning of the tracing history.  Solid lines
   represent entire halo fractions, dashed lines represent MBH
   fractions.  }
    \end{center}
\end{figure}

We explore the evolution of these gas fractions with redshift in
Figure \ref{fig:gasfraction}.  The dashed lines show the instantaneous
fraction of gas mass accreted by the MBH in $hz2$ for each accretion
mode, while the solid lines show the gas fraction of the galaxy as a
whole (colors are as in Figure \ref{fig:BHgrowth}, with the addition
of gray lines representing ``early'' gas which is within the galaxy at
the beginning of the tracing history).  High-$z$ mergers, which occur
in $hz2$'s history around $z \sim 7-10$, have the result of dumping a
large quantity of gas into the host galaxy, which also increases the
fraction the MBH accretes fairly instantaneously.  Such behavior is
evidence of mergers directly fueling MBH growth at high redshift.  On
the other hand, at times when the galaxy's accretion as a whole
becomes more cold-flow-dominated, the MBH also tends to increase the
fraction of cold-flow-originated gas it accretes.  The general trend
is that the modal fraction which the MBH accretes is merely whatever
is available in the galaxy, with a slight delay in time while the
MBH's fuel reservoir adjusts to the composition of the entire system.
(While we show this figure for simulation $hz2$, the results for $hz3$
and $hz4$ are qualitatively similar.)

\begin{figure}[htb]
  \begin{center}
    \includegraphics[scale=0.5]{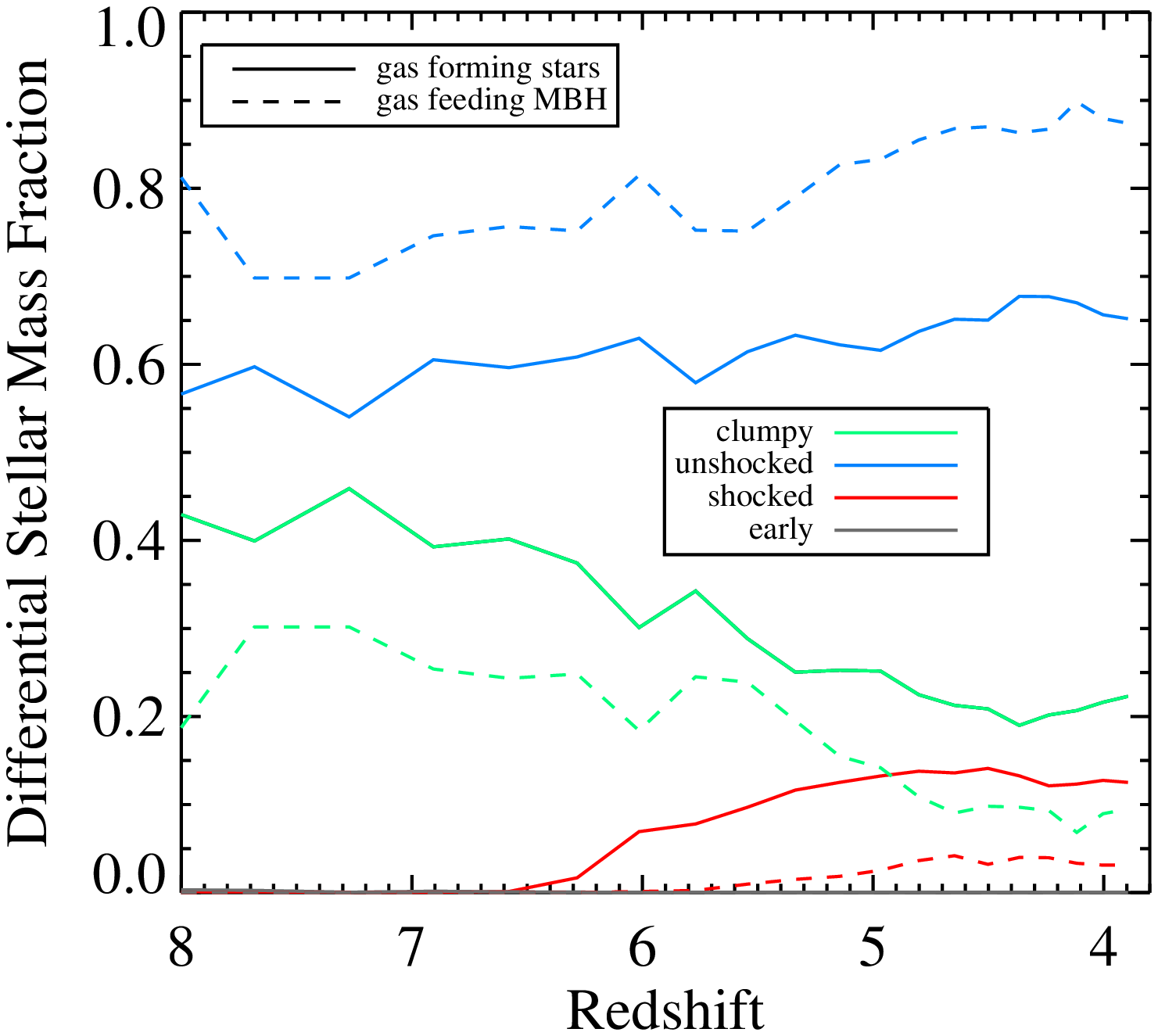}
\caption{ 
   \label{fig:starfraction}  
   Differential stellar mass fraction and MBH accretion fraction vs
   redshift for $hz2$.  Colors are as in Figure \ref{fig:gasfraction}.
   In time bins of $\sim$50 Myr, we determine the accretion origin of
   the gas which has recently formed stars within the central 500
   (physical) parsecs (solid lines) and compare the mass fractions to
   the mass fractions which the MBH accretes (dashed lines) in that
   same time range.  In each case the modal fraction of gas accreted
   by the MBH follows that of the galaxy, in a general sense, though
   the MBH accretes a slightly higher fraction of cold flow gas. }
    \end{center}
\end{figure}

To focus on the galaxy center, which is the residence of the MBH,
in Figure \ref{fig:starfraction} we plot the mass fractions of newly
formed stars within the central 500 parsecs, categorized by the
origins of the gas from which they formed.  For each simulation output
(roughly 50 Myr apart), we determine which stars have formed in the
central region since the previous timestep.  In addition we sum the
gas mass accreted by the MBH during these same timesteps.  These
differential mass fractions of stars and accreted gas show that the
fractions of gas which are accreted by the MBH are slightly different
than the fractions which result in newly formed stars at each
timestep.  The MBH accretes a higher fraction of cold flow gas
throughout its history compared to the composition of the stars which
form at the same time in the same region, and a lower fraction of clumpy and shocked gas.
However, these fractions are still within $10-20$\% of each other at
all times.  Broadly speaking, the origins of accreted gas which
comprise up the overall gaseous, stellar, and MBH makeup in a galaxy
are consistent within this factor. Thus, {\em at any given time, the
  accretion history of the MBH will resemble the accretion history of
  the galaxy; if we can untangle a galaxy's history, we can assume it
  for the MBHs as well.}


\subsection{Angular Momentum}\label{sect:angmom}

\begin{figure}[htb]
  \begin{center}
    \includegraphics[scale=0.5]{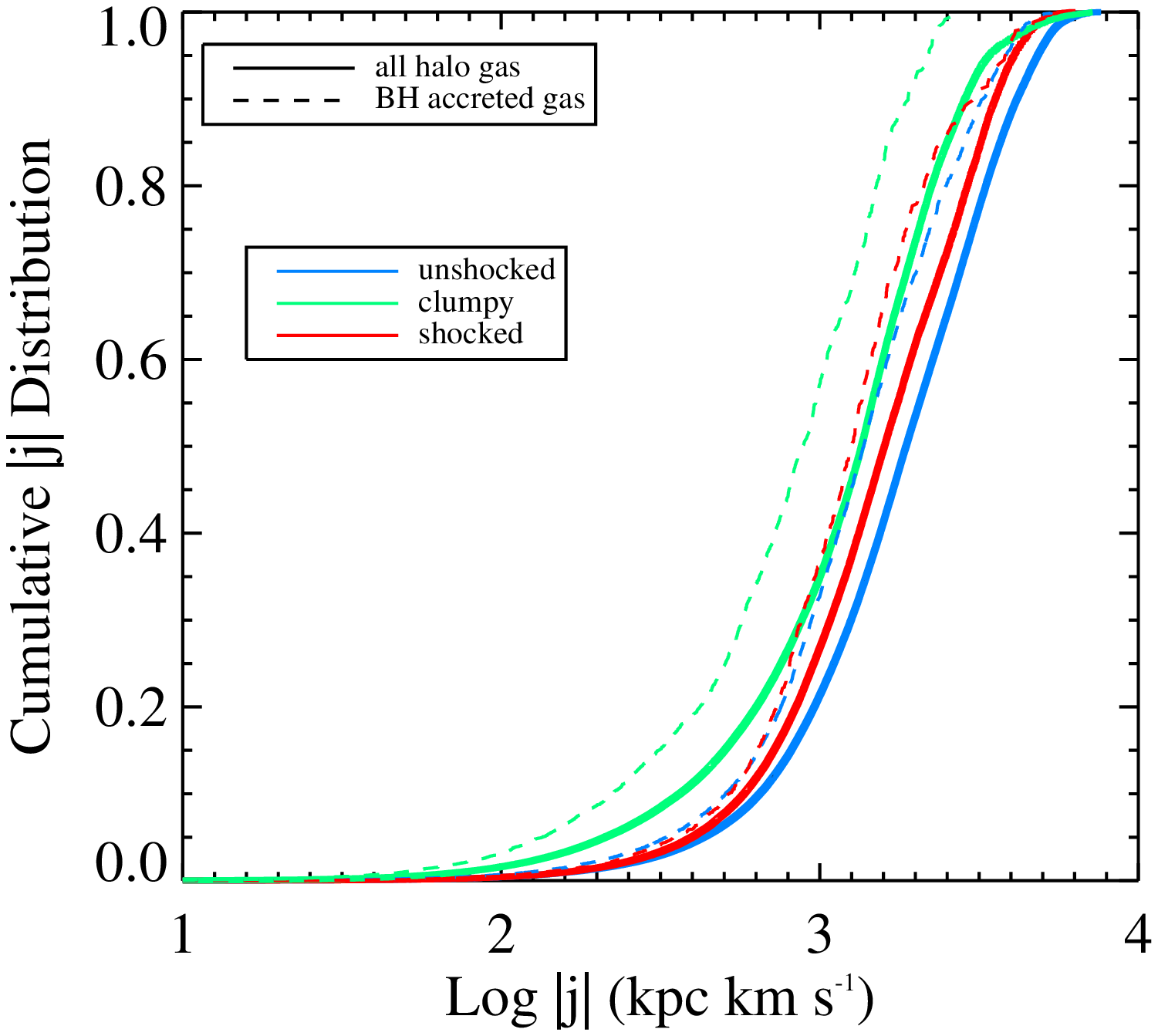}
\caption[Angular Momentum]{ 
   \label{fig:comparej}  
   Normalized cumulative distribution of the log of specific angular
   momentum of gas at the moment it enters the main galaxy in
   simulation $hz2$.  Colors are as in Figure \ref{fig:BHgrowth}.
   Dashed lines represent all of the gas that enters the main galaxy;
   solid lines indicate only that gas which is eventually accreted by
   the central MBH.  For each mode of accretion, the gas which is
   accreted by the MBH has lower angular momentum than the galaxy as a
   whole. }
    \end{center}
\end{figure}

Since the accretion origin of gas does not play a major role in what a
MBH preferentially accretes, one might ask what process {\em does}
play a role in feeding central black holes.  One possibility is the
angular momentum of the inflowing gas; gas with low angular momentum
is more likely to travel directly to the galaxy center instead of
getting caught up in a rotating disk.  To study how the angular
momentum of accreted gas may affect how it is accreted by the MBH, we
have measured this quantity for every gas particle at the moment when
it enters the primary halo.  For smoothly accreted gas, this moment is
when the particle crosses the virial radius; for clumpy gas, the moment
is when the gas can no longer be identified with its host galaxy (at
which point it may have spent several timesteps within $R_{vir}$). 
In Figure \ref{fig:comparej}, we plot the normalized cumulative
distribution of specific angular momenta for each gas particle as it
enters the primary galaxy in simulation $hz2$ (our results are similar
for simulations $hz3$ and $hz4$).  The colored lines represent the
various gas accretion modes onto the halo, as in Figure
\ref{fig:BHgrowth}.  The solid lines represent the entire gas content
of the halo, while the dashed lines are only the gas which is
eventually accreted by the central MBH.  Gas originating in mergers
has markedly less angular momentum overall.  This finding may be a
result of our definition of the moment when gas enters the halo, which
is different for clumpy and smoothly accreted gas (although for
evidence that cold flows carry greater angular momentum see
\citet{Kimm11} and \citet{Stewart13}).  This result is fairly
intuitive - smoothly accreted gas may be torqued by the existing gas
halo on its way to the primary galaxy \citep{Roskar10,Dubois12,Tillson13}, but
gas entering in satellite halos will be able to continue on the
original trajectory of its satellite and initially avoid these strong
torques.  At early times, these satellites tend to fall in along the
filaments \citep{Benson05}, with fairly radial orbits resulting in
early head-on collisions with the primary.

Gas with lower initial angular momentum will reach the galaxy center
more efficiently and thus be more likely to feed the MBH.  Comparing
the solid and dashed lines, in each case the gas which is accreted by
the MBH has overall lower angular momentum than that of the overall
gas
distribution.
While a loss of angular momentum is required for MBH accretion, there
is not a clear argument that one mode of accretion is more efficient
at losing angular momentum than another.  This result strengthens our
argument that the accretion origin of the gas is not the primary
driver behind how MBHs grow, but rather the incoming angular momentum
determines its likelihood for feeding an MBH.


\section{Summary}

We have performed an extensive study of the origins of gas accreted by
high-$z$ MBHs in full cosmological simulations.  We trace the history
of the gas particles and categorize them in terms of smooth accretion
or accretion in mergers, i.e. ``clumpy.''  The smoothly accreted gas
is further divided into shocked and unshocked categories, based on
their temperature and entropy histories as they enter the primary
halo.  We simulate three massive galaxy progenitors to redshift $z =
4$.
We study the gas accretion history for these individual
galaxies and their primary MBHs to determine which gas is accreted
most efficiently and eventually fuels low-luminosity AGNs.

While our MBHs are predominantly fed by cold flows, this result is an
effect of the composition of the galaxy and not because cold flows are
a preferred method of accretion.  MBHs are not picky; they will
accrete whatever gas their host galaxy provides.  The previous study
by \citet{DiMatteo12} reports several MBHs in a simulated volume
growing by accretion from cold flows; however, this gas was identified
by temperature only, and not whether it was accreted smoothly or in
mergers.  The authors report that few major mergers take place; by
default, the host galaxy's gas must consist primarily of smoothly
accreted gas.  The MBHs in their work as well as our own grow via cold
flows because they exist in a galaxy which is fed by cold flows, not
necessarily because cold flows preferentially feed MBHs.

We find that angular momentum is a more fundamental factor in
determining the origins of gas which grows MBHs.  Our findings agree
with those of \citet{Dubois12}, who report efficient feeding of
central MBHs via low angular momentum cold streams.  In addition, they
find that larger mass halos have an increased fraction of radially
infalling gas.  A broader picture is emerging; more massive halos
(which form earlier than low-mass halos, \citep[e.g.][]{Bower06})
exhibit large amounts of low angular momentum gas, which can
efficiently fuel bright quasars and star formation at high redshift \citep[see also][]{Dubois13}.
On the other hand, the angular momentum content of later-collapsing
gas is higher, since it is torqued for a longer period of time.
Additionally, lower mass galaxies eject low angular momentum gas via
feedback processes more efficiently \citep{Governato10,Brook11} due to
their shallower potential wells.  Therefore moderate-mass halos host
lower luminosity AGN, fueled less efficiently because the available
gas has too much inherent spin.

The evidence for merger activity fueling MBH growth is plentiful, but
does not present a full picture.  While bright quasars are certainly
present in the remnants of gas-rich major mergers, we have shown that
MBH fueling at high redshift can occur in a number of ways, and
reflects the gas accretion history of the host galaxy.  Even so, we
have just begun to explore the full parameter space of galaxy mass,
redshift, gas content, and merger history, and how these quantities
affect MBH growth.
Additionally, the journey of the gas from the outer edge of the halo
to the central MBH is still enigmatic.  Torques from mergers and
secular processes drive gas inwards, fueling the black hole, but how
and when this occurs will be the focus of further study.

\acknowledgements

Simulations were run using computer resources and technical support
from NAS.  JB acknowledges support from NSF CAREER award AST-0847696,
as well as support from the Aspen Center for Physics.  MV acknowledges
funding support from NASA, through Award Number ATP NNX10AC84G; from
SAO, through Award Number TM1-12007X, from NSF, through Award Number
AST 1107675, and from a Marie Curie Career Integration grant
(PCIG10-GA-2011-303609).  FG acknowledges support from a NSF grant
AST-0607819 and NASA ATP NNX08AG84G.  JB and TQ acknowledge support
from NASA Grant NNX07AH03G.  AB acknowledges support from The Grainger
Foundation.  The authors thank Kelly Holley-Bockelmann, Ferah
Munshi, and the anonymous referee for their helpful comments.


\end{document}